\documentclass[sigconf, nonacm]{acmart}
\usepackage{multirow}
\usepackage{enumerate}
\usepackage{amsmath}

\usepackage{amssymb}
\usepackage{multirow}
\usepackage{enumerate}
\usepackage{cases}
 \usepackage{amsthm}
 \usepackage{framed}

\usepackage{mathrsfs}

\makeatletter

\newcommand {\Rmnum} [1] {\expandafter \@slowromancap \romannumeral #1@}
\makeatother

\newcommand\vldbpagestyle{plain}
\usepackage{tikz}
\usepackage{bbding}
\usetikzlibrary{matrix,shapes,arrows,positioning,chains, calc,backgrounds}
\usetikzlibrary{decorations.pathreplacing}
\usepackage[misc]{ifsym}
\usepackage{subfigure}
\usepackage{cases}
\usepackage{amsmath}
\usepackage{multirow}
\usepackage{algorithm}
\usepackage{amsmath}
\newtheorem{Theorem}{Theorem}
\newtheorem{Definition}{Definition}

\begin{document}
\title{Quantum Miss-in-the-Middle Attack}
\author{Huiqin Xie$^{1}$, Li Yang$^{2,3}$}
\affiliation{%
  \institution{$^1$Beijing Electronic Science Technology Institute, Beijing {\rm  100070}, China}
  \institution{$^2$State Key Laboratory of Information Security, Institute of Information Engineering, Chinese Academy of Sciences, Beijing {\rm  100093}, China}
  \institution{$^3$Institute of Information Engineering, Chinese Academy of Sciences, Beijing, China}
}
\email{yangli@iie.ac.cn}

\begin{abstract}
Traditional cryptography is facing great challenges with the development of quantum computing. Not only public-key cryptography, the applications of quantum algorithms to symmetric cryptanalysis has also drawn more and more attention. In this paper, we apply quantum algorithms to the miss-in-the-middle technique and propose a quantum algorithm for finding impossible differentials of general block ciphers. We prove that, as long as the attacked block cipher satisfies certain algebraic conditions, the outputs of the quantum algorithm will be impossible differentials of it except for a negligible probability. The proposed quantum algorithm has polynomial quantum complexity and does not require any quantum or classical query to the encryption oracle of the block cipher. Compared with traditional miss-in-the-middle technique, which is difficult to find impossible differentials as the number of rounds increases, the quantum version of miss-in-the-middle technique proposed in this paper is more conducive to find impossible differentials when the block cipher has a large number of rounds.

\end{abstract}

\maketitle

\pagestyle{\vldbpagestyle}
\begingroup
\renewcommand\thefootnote{}\footnote{\noindent
${(\textrm{\Letter})}$ Li Yang \\
~~~~yangli@iie.ac.cn\\
}\addtocounter{footnote}{-1}\endgroup

\section{Introduction}
\label{intro}
The development of quantum computing threatens the security of classical cryptographic schemes. By Shor's algorithm \cite{Sho94}, an adversary with a quantum computer can break any cryptosystem based on discrete logarithm or factorization problems, such as the widely used RSA scheme. Not only public-key cryptography, the research on quantum cryptanalysis of symmetric cryptosystems has also received more and more attention in recent years. A typical example is the quantum algorithm proposed by Grover for searching an unsorted database \cite{Gro96}. Using Grover's algorithm, any exhaustive key search can achieve quadratic speedup. This suggests that, to obtain the ideal security equivalent to that in the classical computing environment, we need to double the key length of symmetric cryptosystems in the post-quantum world.

The brute-force attack only determines the ideal security of cryptosystems. In order to evaluate the actual security of a symmetric cryptographic scheme, it is also required to study other possible attacks that can be executed by quantum adversaries. Simon's algorithm \cite{SD97} plays an important role in the cryptanalysis of symmetric cryptography. For instance, Kuwakado and Morri proposed a quantum distinguisher for 3-round Feistel scheme \cite{KM10} and recovered the key of Even-Mansour scheme \cite{KM12} by Simon's algorithm. Santoli and Schaffiner then extended their result, and proposed a quantum algorithm that executed a forge attack on CBC-MAC scheme based on similar idea \cite{SS17}. In \cite{KL16}, Kaplan et al. also used Simon's algorithm to attack symmetric primitives, including PMAC, GCM, GMAC, CLOC, AEZ and so on. Subsequently, Dong and Wang proposed quantum key-recovery attacks on different rounds of Feistel schemes using the quantum distinguisher presented in \cite{KM10}, and further applied similar method to attack the generalized Feistel schemes \cite{DW18,DW19}. All these attacks are considered under the model of quantum chosen-plaintext attack \cite{BZ13,IJB13,GHS16}, where the adversaries can query the encryption oracle with quantum superpositions of plaintexts. Roetteler and Steinwandt, however, applied Simon's algorithm to the quantum related-key attack model \cite{MR15}, where the adversaries were able to query a superposition of related keys to a block cipher. Afterwards, Hosoyamada and Aoki further studied the quantum related-key attack and used Simon's algorithm to extract the key of two-round Even-Mansour scheme \cite{AK17}.

In addition to constructing specific quantum algorithms for attacking symmetric cryptosystems, applying quantum algorithms to main classical ctyptanalytic tools, such as differential cryptanalysis and linear cryptanalysis, is also significant. Zhou \textit{et al}. applied Grover's algorithm to the key-recovery stage of differential attack and achieved a quadratic speedup \cite{ZLZS15}. Kaplan \textit{et al}. then further used Grover's algorithm in linear attack and different variants of differential attack \cite{KLLNP17}. Afterwards, Xie and Yang applied the Bernstein-Vazirani algorithm \cite{BV97} to the first stage of differential attack, and proposed a quantum algorithm for finding high-probability differentials of block ciphers \cite{XY17}. In this study, we will focus on another important classical analytic tool: miss-in-the-middle attack and improve it by quantum algorithms. The miss-in-the-middle technique was first introduced by Biham, Biryukov and Shamir \cite{BBS99a}. It is a powerful tool for cryptanalysis and has been used to attack many symmetric cryptosystems, such as Skipjack, IDEA, Khufu, CAST-256, and so on \cite{BBS99a,BBS99b}.

\textbf{Our contributions.} In this paper, we further study the application of quantum algorithms to classical ctyptanalytic tools. Specifically, we apply quantum algorithms to the miss-in-the-middle technique and propose a quantum algorithm for finding impossible differentials of general block ciphers. The proposed quantum algorithm has polynomial complexity and has following properties:
\vskip 0.2cm

\noindent
$\bullet\,$ The quantum algorithm does not require any quantum or classical query to the encryption oracle of the block cipher. Compared with many recently proposed quantum attacks \cite{KM10, KM12, SS17, KL16, DW18,DW19}, which require the ability of the attacker to query the encryption oracle with quantum superpositions, our quantum attack is easier to implement in practice.
\vskip 0.2cm

\noindent
$\bullet\,$ As long as the attacked block cipher has impossible differentials that can be connected by two unmatched probability-1 differentials, these impossible differentials must be output by our quantum algorithm. Thus, we can say that, if the classical miss-in-the-middle technique works for the block cipher, our quantum miss-in-the-middle technique must also work for it. Furthermore, we proved that as long as the attacked block cipher satisfies certain algebraic conditions, then except for a negligible probability, all vectors output by the algorithm are impossible differentials of it.
\vskip 0.2cm

\noindent
$\bullet\,$ Miss-in-the-middle technique finds impossible differentials by finding two unmatched probability-1 differentials. In traditional miss-in-the-middle technique, the attacker searches for probability-1 differentials always by searching for probability-1 differential characteristics. Since the probability of differential characteristics generally decrease greatly as the increase of the number of rounds, this will become very difficult when the block cipher has a large number of rounds. In contrast, the quantum version of miss-in-the-middle technique proposed in this paper treats the reduced block cipher as a whole and only cares the input and output differences at the both ends, so is more conducive to find impossible differentials when the block cipher has a large number of rounds.

\section{Preliminaries}
\label{Preliminaries}
In this section, we briefly recall the notations and definitions used in this paper. Suppose $F:\{0,1\}^n\rightarrow\{0,1\}^m$ is a multi-output Boolean function and $m=poly(n)$.
\begin{Definition} A vector $a\in\{0,1\}^n$ is called a linear structure of $F$, if there exists a vector $b\in\{0,1\}^m$ such that
\begin{align}
F(x\oplus a)\oplus F(x)=b,\,\,\,\forall x\in \{0,1\}^n,
\end{align}
where $\oplus$ denotes the bitwise exclusive-or.
\end{Definition}
If $b$ is equal to the $m$-dimensional zero vector, namely $b=0^m$, then the vector $a$ is called a period of $F$. All periods of $F$ constitute a subspace of $n$-dimensional vector space over $\mathbb{F}_2$, which is called period space of $F$. The set of general linear structures of $F$ is not closed under vector addition, so it does not constitute a subspace.

For any $a\in\{0,1\}^n$, $b\in\{0,1\}^m$ that satisfy Eq.(1), the vector $(a,b)$ is called a linear structure pair of $F$. If $(a_1,b_1)$, $(a_2,b_2)$ are two linear structure pairs of $F$, then
$$
F(x\oplus a_1\oplus a_2)\oplus F(x)=F(x\oplus a_1)\oplus b_2\oplus F(x)=b_1\oplus b_2.
$$
That is, $(a_1,b_1)\oplus(a_2,b_2)$ is also a linear structure pair of $F$. Therefore, all linear structure pairs of $F$ constitute a subspace of $n$-dimensional vector space over $\mathbb{F}_2$. We call it linear structure space of $F$ and denote it by $L_F$.

\begin{Definition} A pair of vectors $(a,b)$ ($a\in\{0,1\}^n,b\in\{0,1\}^m$) is called a partial linear structure pair of $F:\{0,1\}^n\rightarrow\{0,1\}^m$, if
\begin{align}
0<\frac{|{x\in\{0,1\}^n|F(x\oplus a)\oplus F(x)=b}|}{2^n}<1.
\end{align}
That is, there must exist at least one vector $x\in\{0,1\}^n$ such that $F(x\oplus a)\oplus F(x)=b$ but it does not hold for all $x\in\{0,1\}^n$.
\end{Definition}

\begin{Definition} A pair of vectors $(a,b)$ ($a\in\{0,1\}^n,b\in\{0,1\}^m$) is called a quasi linear structure pair of $F:\{0,1\}^n\rightarrow\{0,1\}^m$, if the value
\begin{align}
\sigma(n)=\frac{|{x\in\{0,1\}^n|F(x\oplus a)\oplus F(x)=b}|}{2^n}
\end{align}
satisfies $0<\sigma(n)<1$ and $\sigma(n)$ converges to 1 as the parameter $n$ increases. That is, for every $\epsilon>0$, there is an integer $N$ such that $0<1-\sigma(n)<\epsilon$ whenever $n>N$.
\end{Definition}

We can see from the definition that a partial linear structure pair is not a quasi linear structure pair if and only if there exist some constant $p$ such that $\sigma(n)<p<1$ holds for every integer $n$.

Suppose $a\in\{0,1\}^n$, $b\in\{0,1\}^m$ are two vectors. If there exists a vector $x\in\{0,1\}^n$ such that $F(x\oplus a)\oplus F(x)=b$, then we say that $(a,b)$ causes a ``match'' of $F$ at point $x$. $(a,b)$ is a linear structure pair means that it causes matches at all points $x\in\{0,1\}^n$.  $(a,b)$ is a partial linear structure pair means that it causes at least one match but does not cause match at every point. Quasi linear structure pair means that the number of matches it causes converges to 1 as $n$ increases.

Given a Boolean function $F$, a quantum circuit is said to evaluate $F$ if it implements the following unitary operator:
$$
U_F:|x\rangle|y\rangle\rightarrow|x\rangle|y\oplus F(x)\rangle.
$$
Since any quantum circuit can be expressed in terms of gates in some universal, finite set of unitary quantum gates \cite{NC00}, we can assume that the quantum circuit implementing $F$ is composed of gates in an universal gate set. We use the notation $|F|_Q$ to denote the number of universal gates in the quantum circuit implementing $F$. Without loss of generality, we assume that the Hadamard gate $H$ and the controlled--NOT gate $CNOT$ are contained in the universal gate set.

\section{Quantum algorithm for finding linear structures}
In this section we propose a quantum algorithm for finding linear structures of Boolean functions, which in next section will be applied to the miss-in-the-middle attack. We first consider the case in which the Boolean functions has no partial linear structures, then extend this promise for more general situation.

\subsection{Quantum algorithm with no-partial-linear-structure promise}

 Suppose $F:\{0,1\}^n\rightarrow\{0,1\}^m$ is a multi-output Boolean function. $L_F$ is the linear structure space of $F$. Let $t$ denote the dimension of $L_F$. Thus there are $2^t$ vectors in $L_F$. Let $(a_1,b_1)$, $(a_2,b_2)$,$\cdots$, $(a_t,b_t)$ denotes a basis of $L_F$, i.e., $L_F=span{(a_1,b_1),(a_2,b_2),\cdots,(a_t,b_t)}$. $F$ is said to satisfy no-partial-linear-structure (NPLS) promise if, for any vector $(a,b)\notin L_F$, $(a,b)$ cannot cause match at every point $x\in\{0,1\}^n$. That is, if $F(x\oplus a)\oplus F(x)=b$ holds for some $x$, then it must hold that $(a,b)\in L_F$. This promise means $F$ can only have linear structures, no partial linear structures. In this case, we can solve the linear structure space of $F$ by repeatedly executing the following subroutine.
\begin{framed}

\noindent
Subroutine \textbf{FLS}
\vskip 0.2cm

\begin{enumerate}[  1.]
\item Prepare a $(n+2m)$-qubits initial state $|0^n\rangle_{_{\Rmnum{1}}}|0^m\rangle_{_{\Rmnum{2}}}|0^m\rangle_{_{\Rmnum{3}}}$, then perform the Hadamard transform $H^{\otimes (n+m)}$ on the first and second registers, giving
$$
\frac{1}{\sqrt{2^{n+m}}}\sum_{x\in \{0,1\}^n \atop y\in \{0,1\}^m}|x\rangle_{_{\Rmnum{1}}}|y\rangle_{_{\Rmnum{2}}}|0^m\rangle_{_{\Rmnum{3}}}.
$$

\item Execute the unitary operator $U_F$ of $F$ on the first and third registers, giving the state
$$
\frac{1}{\sqrt{2^{n+m}}}\sum_{x\in \{0,1\}^n \atop y\in \{0,1\}^m}|x\rangle_{_{\Rmnum{1}}}|y\rangle_{_{\Rmnum{2}}}|F(x)\rangle_{_{\Rmnum{3}}}.
$$

\item Perform CNOT gates on the second and third registers. Specifically, implement $m$ CNOT gates with the $m$ qubits in the second register  and the $m$ qubits in the third register as the control qubits and the target qubits respectively. The resulting state is
$$
\frac{1}{\sqrt{2^{n+m}}}\sum_{x\in \{0,1\}^n \atop y\in \{0,1\}^m}|x\rangle_{_{\Rmnum{1}}}|y\rangle_{_{\Rmnum{2}}}|F(x)\oplus y\rangle_{_{\Rmnum{3}}}.
$$

\item Measure the third register in the computational basis to get a value $u\in\{0,1\}^{m}$. Then there exist vectors $x_0\in\{0,1\}^n$, $y_0\in\{0,1\}^m$ such that $F(x_0)\oplus y_0=u$.  The first and second registers accordingly collapsed to the state
$$
\frac{1}{\sqrt{2^t}}\sum_{(a,b)\in L_F}|x_0\oplus a\rangle_{_{\Rmnum{1}}}|y_0\oplus b\rangle_{_{\Rmnum{2}}},
$$
where $t$ is the dimension of $L_F$. This will be proved in Lemma 1.

\item Apply the Hadamard transform $H^{\otimes (n+m)}$ to these two registers again, giving:
\begin{align*}
\frac{1}{\sqrt{2^{t+n+m}}}\sum_{(a,b)\in L_F}\sum_{u_1\in \{0,1\}^n \atop u_2\in \{0,1\}^m}(-1)^{(x_0\oplus a)\cdot u_1}(-1)^{(y_0\oplus b)\cdot u_2}&\\
|u_1\rangle_{_{\Rmnum{1}}}|u_2\rangle_{_{\Rmnum{2}}}.&
\end{align*}

\item Measure the above state in the computational basis, giving a vector $u\in\{0,1\}^{n+m}$.
\end{enumerate}
\end{framed}
\vskip 0.1cm

\begin{figure}
  \centering
  \includegraphics[width=8cm]{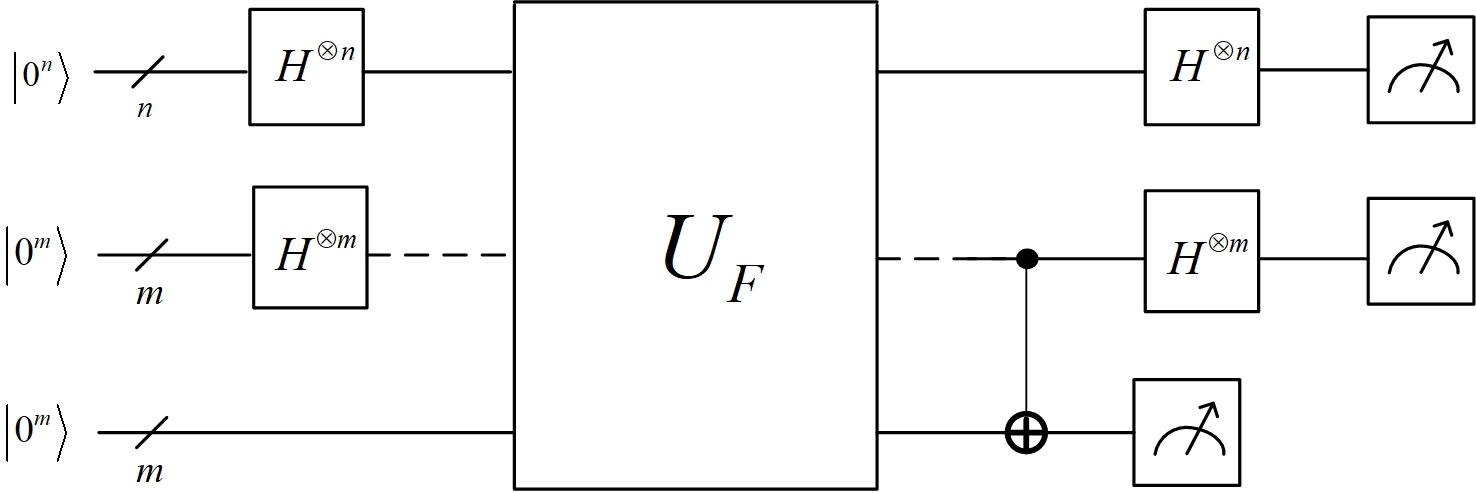}\\
  \caption{Quantum circuit of the subroutine FLS}
\end{figure}

The quantum circuit of subroutine \textbf{FLS} is shown in Figure 1. It will be proved in Lemma 1 that the final measurement of the subroutine \textbf{FLS} always yields a random vector $u$ such that $u\in L_F^{\bot}$. Namely, $u\cdot (a,b)=0$ for every linear structure pair $(a,b)\in L_F$. Repeating the above steps for $O(n+m)$ times, one is expected to obtain $n+m-t$ independent vectors orthogonal to the linear structure space $L_F$, then one can solve $L_F$ efficiently by basic linear algebra.

It is notable that the above subroutine requires that the unitary operator $U_F$ of the function $F$ can be executed. Repeating steps 1-6 once requires $2(n+m)$ single-qubit Hadamard gates, $m$ CNOT gates, and this process needs a total of $n+2m$ qubits.

The following theorem justify the validity of the above method for finding linear structure space.

\begin{Theorem}If $F$ satisfies the NPLS promise, executing steps 1-6 of the subroutine \textbf{FLS} on $F$ always gives a random vector $u\in\{0,1\}^{n+m}$ such that $u\in L_F^{\bot}$. Moreover, the state before the final measurement is exactly an uniform superposition of all vectors orthogonal to $L_F$, i.e.,
$$
\frac{1}{\sqrt{2^{n+m-t}}}\sum_{u\in L_F^{\bot}}(-1)^{u\cdot(x_0,y_0)}|u\rangle.
$$
\end{Theorem}

\noindent
\textbf{Proof}. Suppose the measurement of the step 4 gives a vector $u\in\{0,1\}^m$ and $F(x_0)\oplus y_0=u$. Since $F$ satisfies the NPLS promise, for any $x\in\{0,1\}^n$ and $y\in\{0,1\}^m$,
\begin{align*}
F(x)\oplus y=u &\Leftrightarrow F(x)\oplus y=F(x_0)\oplus y_0\\
&\Leftrightarrow F(x)\oplus F(x_0)=y\oplus y_0\\
&\overset{*}{\Leftrightarrow}(x\oplus x_0,y\oplus y_0)\in L_F.
\end{align*}
The formula $(*)$ holds because $F$ has no partial linear structure pair. Thus,
$$
\{(x,y)|F(x)\oplus y=u\}=\{(x_0\oplus a,y_0\oplus b)|(a,b)\in L_F\}.
$$
Then the state of the first and second register after measurement in step 4 is
$$
\frac{1}{\sqrt{2^t}}\sum_{(a,b)\in L_F}|x_0\oplus a\rangle_{_{\Rmnum{1}}}|y_0\oplus b\rangle_{_{\Rmnum{2}}}.
$$
After performing the Hadamard gates on these two registers in step 5, we obtain the state
\begin{align*}
&\frac{1} {\sqrt{2^{n+m+t}}}\sum_{(a,b)\in L_F, u_1\in \{0,1\}^n \atop u_2\in \{0,1\}^m}(-1)^{(x_0\oplus a)\cdot u_1\oplus(y_0\oplus b)\cdot u_2}|u_1\rangle_{_{\Rmnum{1}}}|u_2\rangle_{_{\Rmnum{2}}}\\
=&\frac{1} {\sqrt{2^{n+m+t}}}\sum_{u_1\in \{0,1\}^n \atop u_2\in \{0,1\}^m}(-1)^{x_0\cdot u_1\oplus y_0\cdot u_2}\sum_{(a,b)\in L_F}(-1)^{(u_1,u_2)\cdot(a,b)}\\
&|u_1\rangle_{_{\Rmnum{1}}}|u_2\rangle_{_{\Rmnum{2}}}\\
=&{\small\frac{1} {\sqrt{2^{n+m+t}}}\sum_{u_1\in \{0,1\}^n \atop u_2\in \{0,1\}^m}(-1)^{x_0\cdot u_1\oplus y_0\cdot u_2}}\\
&{\small\sum_{q_1,q_2,\cdots,q_t\in\{0,1\}}(-1)^{(u_1,u_2)\cdot[q_1(a_1,b_1)\oplus q_2(a_2,b_2)\oplus\cdots\oplus q_t(at,b_t)]}|u_1\rangle|u_2\rangle}.
\end{align*}
The last equation holds since $L_F=span\{(a_1,b_1),(a_2,b_2),\cdots,$ $(a_t,b_t)\}$. This state is actually equal to
\begin{align*}
\frac{1} {\sqrt{2^{n+m+t}}}\sum_{u_1\in \{0,1\}^n \atop u_2\in \{0,1\}^m}(-1)^{x_0\cdot u_1\oplus y_0\cdot u_2}[1+(-1)^{(u_1,u_2)\cdot(a_1,b_1)}]&\\
[1+(-1)^{(u_1,u_2)\cdot(a_2,b_2)}]\cdots[1+(-1)^{(u_1,u_2)\cdot(a_t,b_t)}]|u_1\rangle_{_{\Rmnum{1}}}|u_2\rangle_{_{\Rmnum{2}}}.&
\end{align*}
For any vector $(u_1,u_2)\in\{0,1\}^{n+m}$, if $(u_1,u_2)\cdot(a_i,b_i)\neq0$ for some $i\in\{1,2,\cdots,t\}$, then the amplitude of the state $|u_1,u_2\rangle$ in the above superposition will be zero. In this case $(u_1,u_2)$ will never be a measurement outcome in step 6. Therefore, the final measurement always yields a vector $(u_1,u_2)$ such that $(u_1,u_2)\cdot(a_i,b_i)=0$ for $i=1,2,\cdots,t$. Since $(a_1,b_1),(a_2,b_2),\cdots,(a_t,b_t)$ is a basis of $L_F$, $(u_1,u_2)\in L_F^{\bot}$. Moreover, for any vector $(u_1,u_2)\in L_F^{\bot}$, the amplitude of $|u_1,u_2\rangle$ in the above superposition is equal to
$$
\frac{1}{\sqrt{2^{n+m-t}}}(-1)^{x_0\cdot u_1\oplus y_0\cdot u_2}.
$$
Thus the state before the final measurement is
$$
\frac{1}{\sqrt{2^{n+m-t}}}\sum_{u\in L_F^{\bot}}(-1)^{u\cdot(x_0,y_0)}|u\rangle.
$$
$\hfill{} \Box$

\subsection{Quantum algorithm under weakened condition }

Now we consider the more general situation where $F$ partial linear structure pairs. Namely, there may be vectors $a\in\{0,1\}^n$, $b\in\{0,1\}^m$ such that $(a,b)\notin L_F$ but it causes at least a match of $F$. If the partial linear structure pairs of $F$ cause too many matches, repeating the subroutine \textbf{FLS} may not be able to determine the linear structure space of $F$ efficiently. However, if the number of matches caused by every linear structure pair can be limited properly, this method still works. To illustrate this, we define the following parameter:
\begin{equation}
\delta(F)=\max_{a\in \{0,1\}^n  b\in \{0,1\}^m \atop (a,b)\notin L_F}{\rm Pr}_x[F(x)\oplus F(x\oplus a)=b],
\end{equation}
where
$$
{\rm Pr}_x[F(x)\oplus F(x\oplus a)=b]=\frac{|\{x\in\{0,1\}^n|F(x)\oplus F(x\oplus a)=b\}|}{2^n},
$$
representing the probability that $F(x)\oplus F(x\oplus a)=b$ when $x$ is chosen uniformly from $\{0,1\}^n$. $\delta(F)$ measures how far $F$ is from satisfying the NPLS promise. If $\delta(F)=0$, then $F$ meets the NPLS promise. This parameter also measures the extent to which the linear structure pairs in $L_F$ can be distinguished from the other vectors in $\{0,1\}^{n+m}$. The smaller $\delta(F)$ is, the easier to use the subroutine \text{FLS} to find the linear structure space of $F$.

In the following we show that, as long as $\delta(F)$ does not converge to 1 asymptotically, or in other words, $F$ has no quasi linear structure pair,  repeating the subroutine \textbf{FLS} can still solve the linear structure space of $F$ with a overwhelming probability. We  first present the following algorithm to summarize the process of repeating the subroutine \textbf{FLS} and solving linear equations to find the linear structure space.
\vskip 0.1cm

\begin{framed}

\noindent
Algorithm \textbf{FindStru}
\vskip 0.2cm

\noindent
\hangafter 1
\hangindent 3.4em
\textbf{Input:} $c$ is a constant chosen by the attacker. The quantum oracle access of a Boolean function $F:\{0,1\}^n\rightarrow\{0,1\}^m$ is given. Namely, the unitary operator $U_F$ of $F$ can be performed.

\vskip 0.05cm

\noindent
\textbf{output:} A basis of the linear structure space $L_F$ of $F$.
\vskip 0.05cm

\noindent
\hangafter 1
\hangindent 1.35em
1. Run the subroutine \textbf{FLS} on $F$ for $c(n+m)$ times to get $c(n+m)$ measurement results $u^{(1)},\cdots,u^{(c(n+m))}\in\{0,1\}^{n+m}$.

\noindent
\hangafter 1
\hangindent 1.35em
2. Solve the system of linear equations
$$
\left\{\begin{array}{l}(x,y)\cdot u^{(1)}=0\\
(x,y)\cdot u^{(2)}=0\\
\vdots\\
(x,y)\cdot u^{(c(n+m))}=0,\\
\end{array}
\right.
$$
Suppose $\{(x_1,y_1),(x_2,y_2),\cdots,(x_s,y_s)\}$ is the system of fundamental solutions of above equations, then output the set $\{(x_1,y_1),(x_2,y_2),\cdots,(x_s,y_s)\}$.
\end{framed}
\vskip 0.1cm

The following theorem justify the validity of the algorithm \textbf{FindStru}.
\begin{Theorem}Suppose $F:\{0,1\}^n\rightarrow\{0,1\}^m$ is a multiple-output Boolean function with a linear structure space $L_F$. If running the algorithm \textbf{FindStru} on $F$ with parameter $c$ outputs $\{(x_1,y_1),(x_2,y_2),\cdots,(x_s,y_s)\}$. Let $L=\text{span}\{(x_1,y_1)$ $,(x_2,y_2),\cdots,(x_s,y_s)\}$, then $L_F\subseteq L$. Moreover, if $\delta(F)\leq p_0<1$ for some constant $p_0$, then the probability that $L_F\neq L$ is at most $\big(2(\frac{1+p_0}{2})^c\big)^{n+m}$.
\end{Theorem}


\noindent
\textbf{Proof}. We first prove that $L_F\subseteq L$. Suppose the measurement in step 4 of the subroutine \textbf{FLS} gives a value and $F(x_0)\oplus y_0=u$. Let
$$
G_u=\{(x,y)\in\{0,1\}^n\times\{0,1\}^m|F(x)\oplus y=u\},
$$
Then the measurement in step 4 makes the first and second registers collapse the state
$$
\frac{1}{\sqrt{|G_u|}}\sum_{(x,y)\in\{0,1\}^{n+m}\atop  (x,y)\in G_u}|x\rangle_{_{\Rmnum{1}}}|y\rangle_{_{\Rmnum{2}}}.
$$
Since $L_F=span\{(a_1,b_1),(a_2,b_2),\cdots,(a_t,b_t)\}$, for any $q_1,q_2,$ $\cdots,q_t\in\{0,1\}$, we have
\begin{align*}
&F(x_0\oplus q_1a_1\oplus q_2a_2\oplus\cdots\oplus q_ta_t)\oplus (y_0\oplus q_1b_1\oplus q_2b_2\oplus\cdots\oplus q_tb_t)\\
=&F(x_0)\oplus q_1b_1\oplus q_2b_2\oplus\cdots\oplus q_tb_t\oplus (y_0\oplus q_1b_1\oplus q_2b_2\oplus\cdots\oplus q_tb_t)\\
=&F(x_0)\oplus y_0\\
=&u.
\end{align*}
Therefore,
\begin{align*}
\{(x_0\oplus q_1a_1\oplus q_2a_2\oplus\cdots\oplus q_ta_t,y_0\oplus q_1b_1\oplus q_2b_2\oplus\cdots\oplus q_tb_t)|\\
q_1,q_2,\cdots,q_t\in\{0,1\}\}\subseteq G_u.
\end{align*}
In fact, for any vectors $a\in\{0,1\}^n$, $b\in\{0,1\}^m$,
\begin{align*}
F(x_0\oplus a)\oplus(y_0\oplus b)=u\Leftrightarrow F(x_0)\oplus F(x_0\oplus a)=b\\
\Leftrightarrow (a,b) \text{ causes a match of }F\text{ at point }x_0.
\end{align*}
Since now $F$ does not satisfy the NPLS promise, in addition to the linear structure pairs, there may be other partial linear structure pairs $(\hat{a},\hat{b})$ that cause matches at point $x_0$, thus satisfying $F(x_0\oplus\hat{a})\oplus(y_0\oplus\hat{b})=u$. In this case, for any $q_1,q_2,\cdots,q_t\in\{0,1\}$,
\begin{align*}
F(x_0\oplus\hat{a}\oplus q_1a_1\oplus q_2a_2\oplus\cdots\oplus q_ta_t)\oplus\\
 (y_0\oplus \hat{b}\oplus q_1b_1\oplus q_2b_2\oplus\cdots\oplus q_tb_t)=F(x_0\oplus \hat{a})\oplus(y_0\oplus \hat{b})\\
 =F(x_0)\oplus b\oplus y_0\oplus b=u.
\end{align*}
That is, $(x_0\oplus\hat{a}\oplus\bigoplus_{j=1}^tq_ja_j,y_0\oplus\hat{b}\oplus\bigoplus_{j=1}^tq_jb_j)\in G_u$. For any two unequal partial linear structure pairs $(\hat{a}_1,\hat{b}_1)$ and $(\hat{a}_2,\hat{b}_2)$ that cause matches at point $x_0$, the sets
\begin{align*}
&\{(x_0\oplus\hat{a}_1\oplus \bigoplus_{j=1}^tq_ja_j,y_0\oplus \hat{b}_1\oplus \bigoplus_{j=1}^tq_jb_j)|q_1,q_2,\cdots,q_t\in\{0,1\}\},\\
&\{(x_0\oplus\hat{a}_2\oplus \bigoplus_{j=1}^tq_ja_j,y_0\oplus \hat{b}_2\oplus\bigoplus_{j=1}^tq_jb_j)|q_1,q_2,\cdots,q_t\in\{0,1\}\}
\end{align*}
either have no intersection or are completely equal. Therefore, the measurement of step 4 of the subroutine \text{FLS} causes the first and second registers to collapse into a state in the following form:
\begin{align*}
\frac{1}{\sqrt{2^t(l+1)}}\Big(\sum_{q_1,q_2,\cdots,q_t\in\{0,1\}}|x_0\oplus q_1a_1\oplus q_2a_2\oplus\cdots\oplus q_ta_t\rangle_{_{\Rmnum{1}}}\\
|y_0\oplus q_1b_1\oplus q_2b_2\oplus\cdots\oplus q_tb_t\rangle_{_{\Rmnum{2}}}&\\
+\sum_{q_1,q_2,\cdots,q_t\in\{0,1\}}|x_0\oplus\hat{a}_1 q_1a_1\oplus q_2a_2\oplus\cdots\oplus q_ta_t\rangle_{_{\Rmnum{1}}}\\
|y_0\oplus \hat{b}_1\oplus q_1b_1\oplus q_2b_2\oplus\cdots\oplus q_tb_t\rangle_{_{\Rmnum{2}}}&\\
+\cdots&\\
+\sum_{q_1,q_2,\cdots,q_t\in\{0,1\}}|x_0\oplus\hat{a}_l q_1a_1\oplus q_2a_2\oplus\cdots\oplus q_ta_t\rangle_{_{\Rmnum{1}}}\\
|y_0\oplus \hat{b}_l\oplus q_1b_1\oplus q_2b_2\oplus\cdots\oplus q_tb_t\rangle_{_{\Rmnum{2}}}\Big),&
\end{align*}
where $(\hat{a}_1,\hat{b}_1),(\hat{a}_2,\hat{b}_2),\cdots,(\hat{a}_l,\hat{b}_l)$ are all partial linear structure pairs that cause matches of $F$ at point $x_0$. Let $\hat{a}_0=0^n$, $\hat{b}_0=0^m$, the above formula is equal to
\begin{align*}
\frac{1}{\sqrt{2^t(l+1)}}\sum_{i=0}^l\sum_{q_1,q_2,\cdots,q_t\in\{0,1\}}|x_0\oplus \hat{a}_i\oplus q_1a_1\oplus q_2a_2\oplus\cdots\oplus q_ta_t\rangle_{_{\Rmnum{1}}}\\
|y_0\oplus\hat{b}_i\oplus q_1b_1\oplus q_2b_2\oplus\cdots\oplus q_tb_t\rangle_{_{\Rmnum{2}}}.
\end{align*}
After applying the Hadamard transform $H^{\otimes(n+m)}$ again in the step 5 of the subroutine \textbf{FLS}, the state in the first and second registers is
\begin{align*}
&\frac{1}{\sqrt{2^{n+m+t}(l+1)}}\sum_{u_1\in\{0,1\}^n \atop u_2\in\{0,1\}^m}\sum_{i=0}^l\sum_{q_1,q_2,\cdots,q_t\in\{0,1\}}(-1)^{u_1\cdot x_0+u_2\oplus y_0}\\
&(-1)^{u_1\cdot(\hat{a}_i\oplus q_1a_1\oplus\cdots\oplus q_ta_t)}\cdot(-1)^{u_2\cdot(\hat{b}_i\oplus q_1b_1\oplus\cdots\oplus q_tb_t)}|u_1\rangle_{_{\Rmnum{1}}}|u_2\rangle_{_{\Rmnum{2}}}\\
=&\frac{1}{\sqrt{2^{n+m+t}(l+1)}}\sum_{u_1\in\{0,1\}^n \atop u_2\in\{0,1\}^m}\big(\sum_{i=0}^l(-1)^{u_1\cdot (x_0\oplus\hat{a}_i+u_2\oplus (y_0\oplus\hat{b}_i)}\big)\cdot\\
&[1+(-1)^{(u_1,u_2)\cdot(a_1,b_1)}]\cdot[1+(-1)^{(u_1,u_2)\cdot(a_2,b_2)}]\cdot\cdots\cdot\\
&[1+(-1)^{(u_1,u_2)\cdot(a_t,b_t)}]|u_1\rangle_{_{\Rmnum{1}}}|u_2\rangle_{_{\Rmnum{2}}}.
\end{align*}
Thus, the measurement always yields a random $u$ such that $u\cdot(a_j,b_j)=0$ for $j=1,2,\cdots,t$, which means $(a_1,b_1),\cdots,(a_t,b_t)$ are in the solution space $L$. Therefore, we have $L_F\subseteq L$.

Then we prove that the probability that $L_F\neq L$ is at most $\big(2(\frac{1+p_0}{2})^c\big)^{n+m}$ when $\delta(F)\leq p_0<1$. The probability that $L_F\neq L$ is equal to
\begin{align}
&{\rm Pr}\big[\,\exists \,a\in\{0,1\}^n, b\in\{0,1\}^m \,\,\,\,\text{s.t.} \,\,(a,b)\notin L_F,\,\,u^{(1)}\cdot (a,b)=u^{(2)}\cdot\notag\\
 \notag\\\notag
&(a,b)=\cdots =u^{(c(n+m))}\cdot (a,b)=0\,\big]\notag \\\notag
\leq&\sum_{(a,b)\in\{0,1\}^{n+m}\backslash L_F}{\rm Pr}[u^{(1)}\cdot (a,b)=u^{(2)}\cdot (a,b)=\cdots \\
&=u^{(c(n+m))}\cdot (a,b)=0]\notag\\
&\leq\sum_{(a,b)\in\{0,1\}^{n+m}\backslash L_F}\big(\,{\rm Pr}[u^{(1)}\cdot (a,b)=0]\,\big)^{c(n+m)}\notag\\\notag
\leq&(2^{n+m}-|L_F|)\max_{(a,b)\in\{0,1\}^{n+m}\backslash L_F}\big({\rm Pr}[u^{(1)}\cdot (a,b)=0]\big)^{c(n+m)}\\
\leq&\max_{(a,b)\in\{0,1\}^{n+m}\backslash L_F}\Big(2{\rm Pr}[u^{(1)}\cdot (a,b)=0]^c\Big)^{n+m}
\end{align}
In order to compute ${\rm Pr}[u^{(1)}\cdot (a,b)=0]$, we move the intermediate measurements of the subroutine \textbf{FLS} to its end, then the state before the final measurement can be decomposed into:
\begin{align*}
&\frac{1}{2^{n+m}}\sum_{x\in\{0,1\}^n\atop y\in\{0,1\}^m}\sum_{u_1\in\{0,1\}^n\atop u_2\in\{0,1\}^m}(-1)^{x\cdot u_1+y\cdot u_2}|u_1\rangle|u_2\rangle|F(x)\oplus y\rangle\\
=&\frac{1}{2^{n+m}}\sum_{(u_1,u_2)\in\{0,1\}^{n+m}\atop (u_1,u_2)\cdot(a,b)=0}\sum_{x\in\{0,1\}^n\atop y\in\{0,1\}^m}(-1)^{x\cdot u_1+y\cdot u_2}|u_1\rangle|u_2\rangle|F(x)\oplus y\rangle\\
&+\frac{1}{2^{n+m}}\sum_{(u_1,u_2)\in\{0,1\}^{n+m}\atop (u_1,u_2)\cdot(a,b)=1}\sum_{x\in\{0,1\}^n\atop y\in\{0,1\}^m}(-1)^{x\cdot u_1+y\cdot u_2}|u_1\rangle|u_2\rangle|F(x)\oplus y\rangle.
\end{align*}
Thus the probability that measurement yields $u^{(1)}$ such that $u^{(1)}\cdot (a,b)=0$ is
\begin{align*}
&{\rm Pr}[u^{(1)}\cdot (a,b)=0]\\
=&\|\frac{1}{2^{n+m}}\sum_{(u_1,u_2)\in\{0,1\}^{n+m}\atop (u_1,u_2)\cdot(a,b)=0}\sum_{x\in\{0,1\}^n\atop y\in\{0,1\}^m}(-1)^{x\cdot u_1+y\cdot u_2}|u_1\rangle_{_{\Rmnum{1}}}|u_2\rangle_{_{\Rmnum{2}}}\\
&|F(x)\oplus y\rangle_{_{\Rmnum{3}}}\|^2 \notag\\
=&2^{-2(n+m)}\sum_{(u_1,u_2)\in\{0,1\}^{n+m}\atop (u_1,u_2)\cdot(a,b)=0}\sum_{(x,y)\in\{0,1\}^{n+m}\atop (x',y')\in\{0,1\}^{n+m}}(-1)^{u_1\cdot (z\oplus x')+u_2\cdot(y\oplus y')}\\
&\langle F(x')\oplus y'|F(x)\oplus y\rangle\\
=&2^{-2(n+m)}\sum_{(x,y)\in\{0,1\}^{n+m}\atop (x',y')\in\{0,1\}^{n+m}}\langle F(x')\oplus y'|F(x)\oplus y\rangle\\
&\sum_{(u_1,u_2)\in\{0,1\}^{n+m}\atop (u_1,u_2)\cdot(a,b)=0}(-1)^{u_1\cdot (x\oplus x')+u_2\cdot(y\oplus y')}.
\end{align*}
In order to calculate the above formula, we need to use a conclusion: for any $a\in\{0,1\}^n$, $b\in\{0,1\}^m$,
\begin{equation}
\frac{1}{2^{n+m}}\sum_{(u_1,u_2)\in\{0,1\}^{n+m}\atop (u_1,u_2)\cdot(a,b)=0}(-1)^{u_1\cdot x+u_2\cdot y}=\frac{1}{2}(\delta_{x,0^n}\delta_{y,0^m}+\delta_{x,a}\delta_{y,b}).
\end{equation}
When $(a,b)=(0^n,0^m)$, it is obvious that the formula $(6)$ holds. When $(a,b)\neq(0^n,0^m)$, Let
$$
f(x,y)=\frac{1}{2^{n+m}}\sum_{(u_1,u_2)\in\{0,1\}^{n+m}\atop (u_1,u_2)\cdot(a,b)=0}(-1)^{u_1\cdot x+u_2\cdot y}
$$
By simple calculations it is easy to checked that $f(0^n,0^m)=f(a,b)=\frac{1}{2}$, $\sum{(x,y)\in\{0,1\}^{n+m}}f(x,y)=1$ and $f(x,y)\geq0$. Thus, only when $(x,y)$ is equal to $(0^n,0^m)$ or $(a,b)$, we have $f(x,y)=\frac{1}{2}$. Otherwise, $f(x,y)=0$. This means the formula (5) holds. By the formula (5), the probability that measurement yields $u^{(1)}$ such that $u^{(1)}\cdot (a,b)=0$ is
\begin{align}
&2^{-(n+m)}\sum_{(x,y)\in\{0,1\}^{n+m}\atop (x',y')\in\{0,1\}^{n+m}}\langle F(x')\oplus y'|F(x)\oplus y\rangle\frac{1}{2}(\delta_{x,x'}\delta_{y,y'}+\\
&\delta_{x\oplus a,x'}\delta_{y\oplus b,y'})\notag.\\
=&\frac{1}{2^{n+m+1}}(\sum_{x\in\{0,1\}^{n}\atop y\in\{0,1\}^{m}}\langle F(x)\oplus y|F(x)\oplus y\rangle+\\
&\sum_{x\in\{0,1\}^{n}\atop y\in\{0,1\}^{m}}\langle F(x)\oplus y|F(x\oplus)\oplus y\oplus b\rangle) \notag.\\
=&\frac{1}{2}(1+{\rm Pr}_x[F(x\oplus a)\oplus F(x)=b]).
\end{align}
According to Eq.(5) and Eq.(7), it holds that
\begin{align*}
&{\rm Pr}[L_F\neq L]\\
\leq&\max_{(a,b)\in\{0,1\}^{n+m}\backslash L_F}\Big(2{\rm Pr}[u^{(1)}\cdot (a,b)=0]^c\Big)^{n+m}\\
\leq&\Big(2(\frac{1+\delta(F)}{2})^c\Big)^{n+m}\\
\leq&\Big(2(\frac{1+p_0}{2})^c\Big)^{n+m}.
\end{align*}
$\hfill{} \Box$

By Theorem 1, as long as $c$'s value is at least $\lceil\ln{2}/\ln{\frac{2}{1+p_0}}\rceil$, the solution space $L$ is equal to the linear structure space $L_F$ except for a negligible probability.

\section{Quantum miss-in-the middle attack}

The miss-in-the-middle technique has been widely used in impossible differential cryptanalysis \cite{BBS99a,BBS99b}. Its basic idea is to find two events that propagate from the cipher top and bottom with certainty, but do not match in the middle. This results in an impossible event of the full cipher.

Suppose $E$ is an arbitrary $r$-round block cipher with a blocksize of $n$. Let $F$ denote the first $r-1$ rounds of $E$, and $\mathcal{K}$ be the key space of $F$, namely the subkey space corresponding to the first $r-1$ rounds of $E$. The inputs of $F$ includes a key in $\mathcal{K}$ and a plaintext $x\in\{0,1\}^n$. Fixing a specific key $k\in\mathcal{K}$, the action of $F$ on $x$ is denoted by $F_k(x)$. Suppose $F_k(x)=y$, $F_k(x')=y'$, then $\Delta x=x\oplus x'$ is the input difference and $\Delta y=y\oplus y'$ is the output difference. The pair $(\Delta x,\Delta y)$ is called a differential of $F_k$. If it holds that
$$
F_k(x\oplus \Delta x)\oplus F_k(x)\neq \Delta y, \,\,\forall x\in \{0,1\}^n,
$$
then $(\Delta x,\Delta y)$ is called an impossible differential of $F_k$. When executing an impossible differential cryptanalysis, the attacker first needs to find some impossible differential $(\Delta x,\Delta y)$ of $F_k$, and then uses it to sieve the subkey of the last round of $E$. However, since the key $k$ is private, the attacker cannot access the oracle of $F_k:\{0,1\}^n\rightarrow\{0,1\}^n$. He only knows the function $F:\mathcal{K}\otimes\{0,1\}^n\rightarrow\{0,1\}^n$ without specifying the key. Therefore, in impossible cryptanalysis, the attacker actually looks for a key-independent impossible differential of $F$, which is defined as following:
\begin{Definition}
Suppose $F:\mathcal{K}\otimes\{0,1\}^n\rightarrow\{0,1\}^n$ is a cipher with key space $\mathcal{K}$. $(\Delta x,\Delta y)$ is called a key-independent impossible differential of $F$ if, for any $k\in\mathcal{K}$ and any $x\in\{0,1\}^n$, it holds that
$$
F(k,x\oplus \Delta x)\oplus F(k,x)\neq \Delta y.
$$
\end{Definition}

Key-independent probability-1 differential is defined similarly:
\begin{Definition}
Suppose $F:\mathcal{K}\otimes\{0,1\}^n\rightarrow\{0,1\}^n$ is a cipher with key space $\mathcal{K}$. $(\Delta x,\Delta y)$ is called a key-independent probability-1 differential of $F$ if, for any $k\in\mathcal{K}$ and any $x\in\{0,1\}^n$, it holds that
$$
F(k,x\oplus \Delta x)\oplus F(k,x)= \Delta y.
$$
\end{Definition}
It is obvious that $(\Delta x,\Delta y)$ is a key-independent probability-1 differential of $F$ if only if $(({\bf{0}},\Delta x),\Delta y)$ is a linear structure pair of $F$. Here ${\bf{0}}$ denotes the zero vector in $\mathcal{K}$.

The miss-in-the-middle technique finds key-independent impossible differentials by connecting two unmatched probability-1 differentials. Specifically, for any $v\in\{1,\cdots,r-2\}$, we divide $F$ into two parts: $F=\check{F}_v\cdot\hat{F}_v$, where $\hat{F}_v$ corresponds to the first $v$ rounds of $F$, and $\check{F}_v$ corresponds to the rest $r-1-v$ rounds. The key space $\mathcal{K}$ is accordingly divided into two parts $\mathcal{K}=\mathcal{K}_v^1\otimes\mathcal{K}_v^2$. That is, $\hat{F}_v$ maps $\mathcal{K}_v^1\otimes\{0,1\}^n$ to $\{0,1\}^n$, and $\check{F}_v$ maps $\mathcal{K}_v^2\otimes\{0,1\}^n$ to $\{0,1\}^n$. Fixing a specific key $k\in\mathcal{K}_v^2$, the action of $\check{F}_v$ on $x$ is denoted by $\check{F}_{v,k}(x)$. Define the function
\begin{align*}
\check{F}_v^{(-1)}:\mathcal{K}_v^2\times\{0,1\}^n&\longrightarrow\{0,1\}^n\\
(k\,\,\,,\,\,\,\,\, y)\quad\quad\,\,\,&\longrightarrow (\check{F}_{v,k})^{-1}(y),
\end{align*}
where $(\check{F}_{v,k})^{-1}$ is the inverse of $\check{F}_{v,k}$, namely the decryption function corresponding to the last $r-1-v$ rounds of $F$ given the subkey $k$. If $(\Delta x_1,\Delta y_1)$ and $(\Delta x_2,\Delta y_2)$ are key-independent probability-1 differentials of $\hat{F}_v$ and $\check{F}_v^{(-1)}$, respectively, and $\Delta y_1\neq\Delta y_2$, then $(\Delta x_1,\Delta x_2)$ will be an key-independent impossible differential of $F$ (Figure 2). Therefore, the miss-in-the-middle technique transforms the task of finding key-independent impossible differentials into the task of finding key-independent probability-1 differentials.
\begin{figure}
  \centering
  \includegraphics[width=8cm]{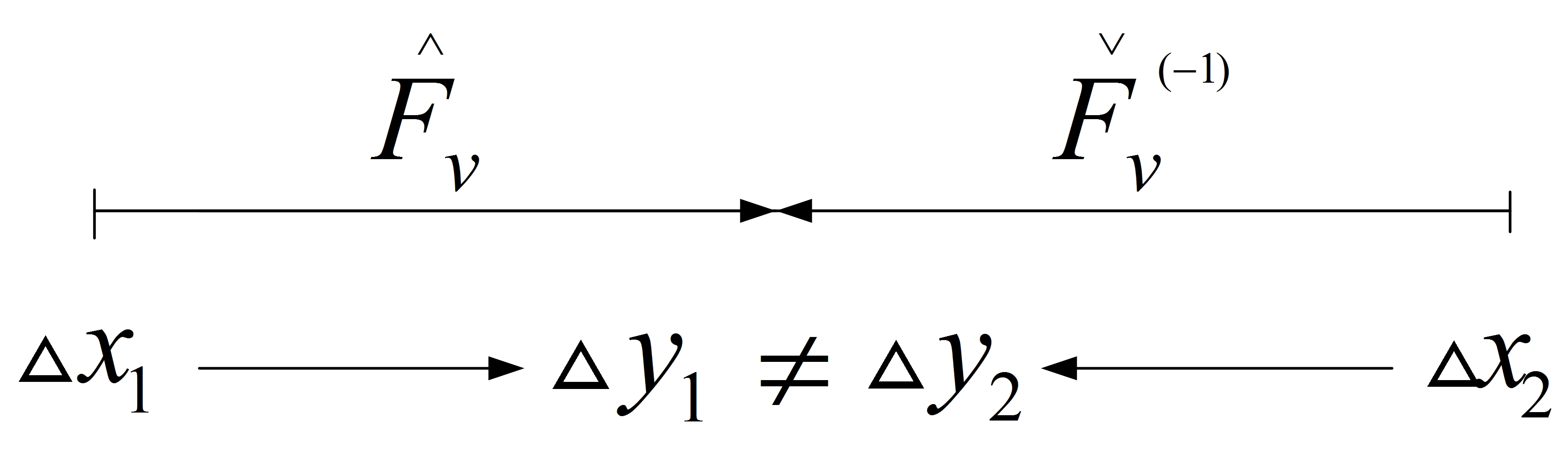}\\
  \caption{Two unmatched key-independent probability-1 differentials resulting in an key-independent impossible differential}
\end{figure}

\subsection{Quantum algorithm for finding key-independent probability-1 differentials}

As mentioned above, to find key-independent impossible differentials of a block cipher by the miss-in-the middle technique, all that remains is to construct a quantum algorithm for finding key-independent probability-1 differentials.

Suppose a cipher $F$ with a key space $\mathcal{K}$ is given. Since a vector $(({\bf{0}},\Delta x),\Delta y)$ is a linear structure pair of $F$ if and only if $(\Delta x,\Delta y)$ is a key-independent probability-1 differential of $F$, we can apply algorithm \textbf{FindStru} to $F$ to find its key-independent probability-1 differentials. The only problem is that we require the linear structure pair $((\Delta k,\Delta x),\Delta y)$ found by the algorithm to satisfy that $\Delta k={\bf{0}}$, where segment $\Delta k$ corresponds to the difference of key bits. To this end, algorithm \textbf{FindStru} needs a minor modification. Suppose $\mathcal{K}=\{0,1\}^m$, the resulting algorithm for finding key-independent probability-1 differentials of a given cipher is as following:
\vskip 0.1cm

\begin{framed}

\noindent
Algorithm \textbf{FindPr1Diff}
\vskip 0.2cm

\noindent
\textbf{Input:} $c$ is a constant chosen by the attacker. $F:\{0,1\}^m\otimes\{0,1\}^n\rightarrow\{0,1\}^n$ is a cipher with key space $\mathcal{K}=\{0,1\}^m$. The access to the unitary operator $U_F$ that computes $F$ quantumly is given.
\vskip 0.05cm

\noindent
\textbf{output:} Key-independent probability-1 differentials of $F$.
\vskip 0.05cm

\noindent
\hangafter 1
\hangindent 1.35em
1. Run the subroutine \textbf{FLS} on $F$ for $c(2n+m)$ times to get $c(2n+m)$ measurement results $u^{(1)},\cdots,u^{(c(2n+m))}\in\{0,1\}^{2n+m}$. Solve the system of linear equations
$$
\left\{\begin{array}{l}(x,y)\cdot (u^{(1)}_{m+1},u^{(1)}_{m+2}\cdots,u^{(1)}_{2n+m})=0\\
(x,y)\cdot (u^{(2)}{m+1},u^{(2)}_{m+2}\cdots,u^{(2)}_{2n+m})=0\\
\vdots\\
(x,y)\cdot (u^{(c(2n+m))}_{m+1},u^{(c(2n+m))}_{m+2}\cdots,u^{(c(2n+m))}_{2n+m})=0,\\
\end{array}
\right.
$$
where the unknowns $x$ and $y$ both have $n$ bites, and $(u^{(j)}_{m+1},u^{(j)}_{m+2}\cdots,u^{(j)}_{2n+m})$ is the last $2n$ bits of the vector $u^{(j)}$ for $j=1,2,\cdots,c(2n+m)$. Suppose $\{(x_1,y_1),(x_2,y_2),\cdots,(x_t,y_t)\}$ is the system of fundamental solutions of above equations, then output the set $\{(x_1,y_1),(x_2,y_2),\cdots,(x_t,y_t)\}$.
\end{framed}
\vskip 0.1cm

Let $L=\text{span}\{(x_1,y_1),(x_2,y_2),\cdots,(x_t,y_t)\}$. The following theorem shows that, except for a negligible probability, $L$ is the set of all key-independent probability-1 differentials of $F$.
\begin{Theorem}Suppose $F:\{0,1\}^m\otimes\{0,1\}^n\rightarrow\{0,1\}^n$ is a block cipher with key space $\mathcal{K}=\{0,1\}^m$. If running algorithm \textbf{FindPr1Diff} on $F$ with a parameter $c$ outputs  $\{(x_1,y_1),(x_2,y_2),\cdots,(x_t,y_t)\}$, and $L={\rm span}\{(x_1,y_1),(x_2,y_2),$ $\cdots,(x_t,y_t)\}$, then $L$ contains all key-independent probability-1 differentials of $F$. Moreover, if $\delta(F)\leq p_0<1$ for some constant $p_0$, then the probability that $L$ contains other vectors that are not key-independent probability-1 differential of $F$ is at most $\big(2(\frac{1+p_0}{2})^c\big)^{2n+m}$.
\end{Theorem}

\noindent
\textbf{Proof}. Note that $(x_i,y_i)$ is a solution of
$$
\left\{\begin{array}{l}(x,y)\cdot (u^{(1)}_{m+1},u^{(1)}_{m+2}\cdots,u^{(1)}_{2n+m})=0\\
(x,y)\cdot (u^{(2)}_{m+1},u^{(2)}_{m+2}\cdots,u^{(2)}_{2n+m})=0\\
\vdots\\
(x,y)\cdot (u^{(c(2n+m))}_{m+1},u^{(c(2n+m))}_{m+2}\cdots,u^{(c(2n+m))}_{2n+m})=0,\\
\end{array}
\right.
$$
is equivalent to that $(0,x_i,y_i)$ is a solution of
$$
\left\{\begin{array}{l}(k,x,y)\cdot u^{(1)}=0\\
(k,x,y)\cdot u^{(2)}=0\\
\vdots\\
(k,x,y)\cdot u^{(c(2n+m))}=0\\
k=0^m.\\
\end{array}
\right.
$$
Therefore, $\{(0^m,x_1,y1),(0^m,x_2,y_2),\cdots,(0^m,x_t,y_t)\}$ can be seen as the output when running algorithm \textbf{FindStru} on $F$, under the additional condition that the first $m$ bits must be zeros. Let
\begin{align*}
L_F^0=\{(\Delta k,\Delta x,\Delta y)|(\Delta k,\Delta x,\Delta y)\\
 \text{is a linear structure pair of } F \,\bigwedge\, \Delta k=0^m\}.
\end{align*}
Then $L_F^0$ is a subspace of the linear structure space $L_F$. Let $L'=\text{span}\{(0^m,x_1,$ $y_1),(0^m,x_2,y_2),\cdots,$ $(0^m,x_t,y_t)\}$. By Theorem 2, $L'\subseteq L_F^0$, and the probability that $L'\neq L_F^0$ is at most $(2(\frac{1+p_0}{2})^c)^{2n+m}$. Since $(\Delta x,\Delta y)$ is a key-independent probability-1 differential of $F$ if only if $((0^m,\Delta x),\Delta y)$ is a linear structure pair of $F$, we have that $L$ contains all key-independent probability-1 differentials of $F$, and the probability that $L$ contains other vectors that are not key-independent probability-1 differential of $F$ is at most $\big(2(\frac{1+p_0}{2})^c\big)^{2n+m}$.

$\hfill{} \Box$

\subsection{Quantum algorithm for finding key-independent impossible differentials}

In order to find key-independent impossible differentials of a block cipher $F$, for each $v\in\{1,\cdots,r-2\}$, we divide $F$ into two parts: $F=\check{F}_v\cdot\hat{F}_v$, with subkey space $\mathcal{K}_v^1$ and $\mathcal{K}_v^2$ respectively. As mentioned earlier, if $(\Delta x_1,\Delta y_1)$ and $(\Delta x_2,\Delta y_2)$ are key-independent probability-1 differentials of $\hat{F}_v$ and $\check{F}_v^{(-1)}$ respectively, and $\Delta y_1\neq\Delta y_2$, then $(\Delta x_1,\Delta x_2)$ will be a key-independent impossible differential of $F$. Therefore, by applying algorithm \textbf{FindPr1Diff} on $\hat{F}_v$ and $\check{F}_v^{(-1)}$ to obtain key-independent probability-1 differentials of them respectively, one is able to find key-independent impossible differentials of $F$.

Given a cipher $F$ with key space $\mathcal{K}=\{0,1\}^m$, the algorithm for finding key-independent impossible differentials of $F$ is as following:
\vskip 0.1cm

\begin{framed}

\noindent
Algorithm \textbf{FindImDiff}
\vskip 0.2cm

\noindent
\textbf{Input:} $c$ is a constant chosen by the attacker. $F:\{0,1\}^m\otimes\{0,1\}^n\rightarrow\{0,1\}^n$ is a cipher with key space $\mathcal{K}=\{0,1\}^m$.
\vskip 0.05cm

\noindent
\textbf{output:} Key-independent impossible differentials of $F$.
\vskip 0.05cm

\noindent
\hangafter 1
\hangindent 1.35em
1. For $v=1,2,\cdots, r-2$, divide $F$ into two parts: $F=\check{F}_v\cdot\hat{F}_v$ as described above. Since the construction of $F$ is public, the unitary operators $U_{\check{F}_v}$ and $U_{\hat{F}_v^{(-1)}}$ that compute $\check{F}_v$ and $\hat{F}_v^{(-1)}$ respectively is accessible. For each $v=1,2,\cdots, r-2$, do the following:

\hangafter 1
\hangindent 4em
\quad1-1.Apply algorithm \textbf{FindPr1Diff} to the cipher $\hat{F}_v$ with key space $\mathcal{K}_v^1$ and parameter $c$, obtaining the set $A_v$.

\hangafter 1
\hangindent 4em
\quad1-2.Apply algorithm \textbf{FindPr1Diff} to the cipher $\check{F}_v^{(-1)}$ with key space $\mathcal{K}_v^2$ and parameter $c$, obtaining the set $B_v$.

\noindent
\hangafter 1
\hangindent 1.35em
2.Initialize set H to an empty set. For $v=1,2,\cdots, r-2$, do the following:

\quad For any $(\Delta x_1,\Delta y_1)\in \text{{\rm span}}A_v$, any $(\Delta x_2,\Delta y_2)\in \text{{\rm span}}B_v$,

\quad if $\Delta x_1\neq0$, $\Delta x_2\neq0$ and $\Delta y_1\neq \Delta y_2$,

\quad let $H=H\cup\{(\Delta x_1,\Delta x_2)\}$.

\noindent
\hangafter 1
\hangindent 1.35em
3.Output the set $H$.
\end{framed}
\vskip 0.1cm

Note that, since the construction of cipher $F$ is public, the attacker can construct the unitary operators of all $\check{F}_v$ and $\hat{F}_v$ for $v=1,2,\cdots,r-2$ by himself. Therefore, algorithm \textbf{FindImDiff} does not require any query to the encryption oracle of the block cipher, no matter quantum or classical. (Although the attacker does need classical queries to the encryption oracle for the key-recovery phase in the impossible differential cryptanalysis.) Compared with many quantum attacks that have been proposed \cite{KM10, KM12, SS17, KL16, DW18,DW19}, our attack does not require the ability of the attacker to make quantum chosen-plaintext queries \cite{IJB13,GHS16}, thus is easier to implement.

\subsection{Analysis of the algorithm }

In this subsection, we analyze the validity and complexity of algorithm \textbf{FindImDiff}. To do this, we first define the parameter
$$
\hat{\delta}(F)=\max\{\delta(\hat{F}_v),\delta(\check{F}_v^{(-1)}):1\leq v\leq r-2 \},
$$
where $\delta(\hat{F}_v),\delta(\check{F}_v^{(-1)})$ are defined as in Eq.$(4)$. For any $v\in\{1,2,\cdots,r-2\}$, $\hat{F}_v$ and $\check{F}_v^{(-1)}$ are both reduced version of the block cipher $E$, with $v$ and $r-1-v$ rounds respectively. According to Eq.$(4)$, the parameter $\delta(\hat{F}_v)$ measures the distance between the linear structure pairs of $\hat{F}_v$ with the other vectors. Thus, $\hat{\delta}(F)$ quantifies the extent to which the linear structure pairs of the reduced versions of $E$ can be distinguished from the other vectors. The following theorem can be derived straightforward from Theorem 3.
\begin{Theorem}Suppose $F:\{0,1\}^m\otimes\{0,1\}^n\rightarrow\{0,1\}^n$ is a cipher with key space $\mathcal{K}=\{0,1\}^m$, and $\hat{\delta}(F)\leq p_0<1$ for some constant $p_0$. If running algorithm \textbf{FindImDiff} on $F$ with parameter $c$ outputs a set $H$, then all vectors in $H$ are key-independent impossible differentials of $F$. Moreover, the probability that $H$ contains the vectors that are not key-independent impossible differentials of $F$ is at most $2\big(2(\frac{1+p_0}{2})^c\big)^{2n}$.
\end{Theorem}

The condition that $\hat{\delta}(F)\leq p_0<1$ can be understood as requiring all vectors that are not linear structure pairs of any reduced version of $E$ not to be close to ``being a linear structure pair of a reduced version of it''. Theorem 4 shows that, under this condition, all vectors in $H$ are key-independent impossible differentials of $F$ except for a negligible probability. In fact, from Theorem 3 we can draw a further conclusion: as long as $F$ has a key-independent impossible differential that can be connected by two unmatched probability-1 differentials, it must be output by the algorithm \textbf{FindImDiff}. This holds no matter whether $\hat{\delta}(F)$ is bounded by a constant less than 1.

To explain this, suppose $F$ has a key-independent impossible $(\Delta x_1,\Delta x_2)$ that can be connected by two unmatched key-independent probability-1 differentials. Then there exist $\Delta y_1, \Delta y_2$ such that $\Delta y_1\neq\Delta y_1$, and $(\Delta x_1,\Delta y_1)$ and $(\Delta x_2, \Delta y_2)$ are key-independent probability-1 differentials of $\hat{F}_v$ and $\check{F}_v^{(-1)}$ respectively for some $v\in\{1,2,\cdots, r-2\}$. According to Theorem 3, $(\Delta x_1,\Delta y_1)$ must be in the set $\text{{\rm span}}A_v$, and $(\Delta x_2,\Delta y_2)$ must be in the set $\text{{\rm span}}B_v$. Thus, $(\Delta x_1,\Delta x_2)$ must be in the set $H$ output by algorithm \textbf{FindImDiff}. This means that any nontrivial key-independent impossible differential of $F$ that are connected by two unmatched probability-1 differentials can be found by algorithm \textbf{FindImDiff}. Therefore, in some degree, we can say that our quantum miss-in-the-middle technique always works for the block cipher as long as the classical miss-in-the-middle technique works for it.

In order to show the advantages of algorithm \textbf{FindImDiff}, we compare it with traditional miss-in-the-middle technique. The miss-in-the-middle technique requires to find two unmatched probability-1 differentials. In classical cryptanalysis, the attacker looks for probability-1 differentials always by searching for probability-1 differential characteristics. However, since the probability of differential characteristics generally decrease greatly as the increase of the number of rounds, finding probability-1 differential characteristics becomes more and more difficult when the number of rounds increases. This is also the reason why in traditional differential analysis, it is difficult to find high-probability or impossible differentials when the number of rounds is large. By contrast, in our quantum attack, algorithm \textbf{FindImDiff} treats $\hat{F}_v$ and $\check{F}_v^{(-1)}$ as a whole and cares only the input and output differences at their both ends, instead of specific differential characteristics, so it suffers from much smaller effect of the increase of the number of rounds. Therefore, compared with the conventional miss-in-the-middle technique, algorithm \textbf{FindImDiff} is more conducive to find impossible differentials when the block cipher has a large number of rounds.

Next we analyze the complexity of algorithm \textbf{FindImDiff}. Algorithm \textbf{FindImDiff} does not require any query to the encryption oracle of the block cipher, no matter quantum or classical. We evaluate its complexity from three perspectives: the amount of universal gates, the number of qubits required and the amount of calculations needed for classical computing part of the algorithm.

To evaluate the amount of universal gates, suppose $\mathcal{K}_1^v=\{0,1\}^{l_v}$, $\mathcal{K}_2^v=\{0,1\}^{h_v}$, where $l_v+h_v=m$. Running algorithm \textbf{FindPr1Diff} on $\hat{F}_v$ requires calling the subroutine \textbf{FLS} for $c(2n+l_v)$ times, and each time needs $2(2n+l_v)$ Hadamard gates, $n$ single-qubit CNOT gates and $|\hat{F}_v|_Q$ additional universal gates. Likewise, running algorithm \textbf{FindPr1Diff} on $\check{F}_v^{(-1)}$ calls the subroutine \textbf{FLS}e for $c(2n+h_v)$ times, and each time needs a total of $2(2n+h_v)+n+|\check{F}_v^{(-1)}|_Q$ universal gates. Therefore, the total number of quantum universal gates required by algorithm \textbf{FindImDiff} is
\begin{align*}
&\sum_{v=1}^{r-2}\Big\{c(2n+l_v)\big[2(2n+l_v)+n+|\hat{F}_v|_Q\big]+c(2n+h_v)\\
&\big[2(2n+h_v)+n+|\check{F}_v^{(-1)}|_Q\big]\Big\}\\
\leq&c(2n+m)\sum_{v=1}^{r-2}\big[2(l_v+h_v)+10n+(|\hat{F}_v|_Q+|\check{F}_v^{(-1)}|_Q)\big]\\
{\small\overset{(*)}{=}}&10c(r-2)(2n+m)n+c(2n+m)\sum_{v=1}^{r-2}\big(2m+|F|_Q\big)\\
=&c(r-2)(2n+m)(10n+2m+|F|_Q)\in poly(n),
\end{align*}
where the formula $(*)$ holds because $l_v+h_v=m$ and $$|\hat{F}_v|_Q+|\check{F}_v^{(-1)}|_Q=|\hat{F}_v|_Q+|\check{F}_v|_Q=|F|_Q.$$
Since $F$ is the encryption function corresponding the first $r-1$ rounds of the block cipher $E$, it can be efficiently implemented by a quantum circuit, so $|F|_Q$ is a polynomial of $n$. Therefore, the total number of quantum universal gates required by algorithm \textbf{FindImDiff} is a polynomial of security parameter $n$.

Running algorithm \textbf{FindPr1Diff} on $\hat{F}^{(v)}$ needs $2n+l_v+n=3n+l_v$ qubits. Running algorithm \textbf{FindPr1Diff} on $(\check{F}^{(v)})^{-1}$ needs $2n+h_v+n=3n+h_v$ qubits. Since $l_v,h_v\leq m$ and the qubits can be reused, $3n+m$ qubits are sufficient for executing algorithm \textbf{FindImDiff}.

Classical computing part of algorithm \textbf{FindImDiff} includes two phases: solving a system of linear equations each time algorithm \textbf{FindPr1Diff} is called; computing the set $H$. For the second phase, since $\hat{F}_v$, $\check{F}_v$ are both reduced versions of the block cipher $E$, they have very few key-independent probability-1 differentials or linear structures. Let $\alpha=\max_v\{|{\rm span}A_v|,|{\rm span}B_v|\}$. $\alpha$ can be viewed a small number. The amount of calculations needed for computing the set $H$ is $O((r-2)\alpha^2n)$. As for the part of solving systems of linear equations, running algorithm \textbf{FindPr1Diff} on $\hat{F}_v$ needs to solve a system of linear equations with $c(2n+l_v)$ equations and $2n+l_v$ variables. This needs $O(c(2n+l_v)^3)$ calculations. Likewise, running algorithm \textbf{FindPr1Diff} on $\check{F}_v^{(-1)}$ needs $O(c(2n+h_v)^3)$ calculations. Therefore, the amount of calculations needed for solving linear systems is
\begin{align*}
&\sum_{v=1}^{r-2}\big[c(2n+l_v)^3+c(2n+h_v)^3\big]\\
\leq&c\sum_{v=1}^{r-2}(4n+l_v+h_v)^3\\
=&c(r-2)(4n+m)^3.
\end{align*}
The total amount of calculations needed for classical computing part is $O((r-2)[c(4n+m)^3+\alpha^2n])$, where $\alpha$ is a small number.
\section{Conclusion}

In this paper, we apply quantum algorithms to the miss-in-the-middle technique and propose a quantum algorithm for finding impossible differentials of general block ciphers. Our algorithm does not require any quantum or classical query to the encryption oracle of the block cipher, and the amount of qubits and universal quantum gates required for executing the algorithm is polynomial.

To justify the validity of the proposed quantum algorithm, we demonstrated that, as long as the block cipher has key-independent impossible differentials that can be connected by two unmatched probability-1 differentials, these impossible differentials will always be output by the algorithm. Therefore, in some degree, we can say that our quantum miss-in-the-middle technique always works for the block cipher as long as the classical miss-in-the-middle technique works for it. Furthermore, we proved that if the reduced version $F$ of the block cipher with $r-1$ rounds satisfies that $\hat{\delta}(F)\leq p_0<1$ for some constant $p_0$, then except for a negligible probability, all vectors output by the algorithm are key-independent impossible differentials of $F$. Compared with traditional miss-in-the-middle technique, in which it becomes more difficult to find impossible differentials as the number of rounds increases, the quantum version of miss-in-the-middle technique proposed in this paper is more conducive to find impossible differentials when the block cipher has a large number of rounds.

\section*{Acknowledgement}
This work was supported by National Natural Science Foundation of China (Grant No.61672517) and National Cryptography Development Fund (Grant No. MMJJ201 70108).

\end{document}